\documentclass[aps,twocolumn,pra,superscriptaddress,amsmath,showpacs,tightenlines,pdflatex,longbibliography]{revtex4-1}
\usepackage{amssymb}
\usepackage{amsmath}
\usepackage{dcolumn}
\usepackage{multirow}
\usepackage{graphicx}
\usepackage{mathrsfs}
\usepackage{appendix}
\usepackage{graphicx}
\usepackage{booktabs}
\usepackage{color}

\def \new#1{#1}

\newcommand{\ket}[1]{\mbox{$|#1\rangle$}}

\usepackage{url}
\usepackage[colorlinks]{hyperref}
\hypersetup{%
    plainpages=true,
    breaklinks=true,
    hypertexnames=false,
    pageanchor=true,
    colorlinks=true,
    linkcolor={blue},
    citecolor={red},
    urlcolor={blue},
    anchorcolor={black}
}

\begin{document}

\title{Heralded quantum controlled phase gates with dissipative \\ dynamics in macroscopically-distant resonators}

\author{Wei Qin}
\affiliation{CEMS, RIKEN, Wako-shi, Saitama 351-0198, Japan}
\affiliation{School of Physics, Beijing Institute of Technology,
Beijing 100081, China}

\author{Xin Wang}
\affiliation{CEMS, RIKEN, Wako-shi, Saitama 351-0198, Japan}
\affiliation{Institute of Quantum Optics and Quantum Information,
School of Science, Xi'an Jiaotong University, Xi'an 710049, China}

\author{Adam Miranowicz}
\affiliation{CEMS, RIKEN, Wako-shi, Saitama 351-0198, Japan}
\affiliation{Faculty of Physics, Adam Mickiewicz University,
61-614 Pozna\'n, Poland}

\author{Zhirong Zhong}
\affiliation{CEMS, RIKEN, Wako-shi, Saitama 351-0198, Japan}
\affiliation{Department of Physics, Fuzhou University, Fuzhou
350002, China}

\author{Franco Nori}
\affiliation{CEMS, RIKEN, Wako-shi, Saitama 351-0198, Japan}
\affiliation{Physics Department, The University of Michigan, Ann
Arbor, Michigan 48109-1040, USA}

\begin{abstract}
Heralded near-deterministic multi-qubit controlled phase gates
with integrated error detection have recently been proposed by
Borregaard {\it et al}. [Phys. Rev. Lett. {\bf 114}, 110502
(2015)]. This protocol is based on a single four-level atom (a
heralding quartit) and $N$ three-level atoms (operational qutrits)
coupled to a single-resonator mode acting as a cavity bus. Here we
generalize this method for two distant resonators without the
cavity bus between the heralding and operational atoms.
Specifically, we analyze the two-qubit controlled-Z gate and its
multi-qubit-controlled generalization (i.e., a Toffoli-like gate)
acting on the two-lowest levels of $N$ qutrits inside one
resonator, with their successful actions being heralded by an
auxiliary microwave-driven quartit inside the other resonator.
Moreover, we propose a circuit-quantum-electrodynamics realization
of the protocol with flux and phase qudits in linearly-coupled
transmission-line resonators with dissipation. These methods offer
a quadratic fidelity improvement compared to cavity-assisted
deterministic gates.
\end{abstract}
 \maketitle
\section{introduction}
\label{se:introduction}

The ability to carry out quantum gates is one of the central
requirements for a functional quantum computer~\cite{chuang}. For
this reason, quantum gates have been extensively studied in
theoretical and experimental in various systems. These include
(for reviews see~\cite{rBuluta11,rXiang13} and references
therein): superconducting circuits, trapped ions, diamond
nitrogen-vacancy centers, semiconductor nanostructures, or
linear-optical setups. Despite of such substantial efforts, the
environment-induced decoherence still represents a major hurdle in
the quest for perfect quantum gates. To protect quantum systems
from decoherence, three basic approaches have been explored;
namely, quantum error correction, dynamical decoupling, and
decoherence-free subspaces (for reviews
see~\cite{Lidar03,Devitt13} and references therein). Such
approaches try to cope with decoherence, and thus, to prevent the
leakage of quantum information from a quantum system into its
environment.

Alternatively, there exists a very distinct way, where decoherence
acts as a resource, rather than as a traditional noise
source~\cite{Bose99,Chimczak05,Dresource1,Dresource2,Dresource3}.
Recent progress in treating open quantum systems has yielded an
effective operator formalism~\cite{Dentanglement1,effmaseq}.
As in Ref.~\cite{effmaseq}, (i) when the interactions between the ground-
and excited-state subspaces of an open system initially in its
ground state are sufficiently weak (so it can be considered as
perturbations of the two subspaces) and also (ii) when the
interactions inside the ground-state subspace are much smaller
than those inside the excited-state subspace, one can
adiabatically eliminate these excited states in the presence of
both unitary and dissipative dynamics, and obtain an effective
master equation containing the effective Hamiltonian, as well as
the effective Lindblad operators, associated only with the ground
states. In addition to reducing the complexity of the time
evolution for an open quantum system, this approximation treatment
is applicable to the explorations of decay processes, hence may
lead to a better performance than that in the case of relying upon
coherent-unitary dynamics.

So far, such a formalism has been widely used for dissipative
entanglement preparation~\cite{Dentanglement1,Dentanglement2,
Dentanglement3,Dentanglement4, Dentanglement5,
Dentanglement6,Dentanglement7}, quantum phase
estimation~\cite{Dphase} and other
applications~\cite{other1,other2,other3,other4}. In particular,
Borregaard {\it et al.} presented a heralded near-deterministic
method for quantum gates in a single optical cavity~\cite{Dgate1},
with a significant improvement in the error scaling, compared to
deterministic cavity-based gates. However, in order to carry out
scalable quantum information processing, a distant herald for
quantum gates in coupled resonators is of a great importance
concerning both experimental feasibility and fundamental tests of
quantum mechanics.

The goal of this paper is to propose and analyze an approach to
heralding controlled quantum gates in two macroscopically distant
resonators, by generalizing the single-resonator method of
Borregaard {\it et al.}~\cite{Dgate1}. We use an auxiliary quartit
atom, which is located in one resonator and driven by two coherent
fields, to distantly herald controlled phase gates acting on the
two lowest levels of qutrit atoms, which are located in the other
resonator. These controlled phase operations studied here include
the two-qubit controlled-Z gate (controlled-sign gate) and its
multi-qubit-controlled version referred to as a Toffoli-like gate.
We also propose a realization of these gates using superconducting
artificial atoms in a dissipative circuit quantum electrodynamics
(QED) system.

Circuit-QED systems, which are composed of superconducting
artificial atoms coupled to superconducting resonators, offer
promising platforms for quantum engineering and
quantum-information processing~\cite{rYou05,rClarke08,rDiCarlo10,
rBuluta11,rLucero12,rGeorgescu14}. Although artificial and natural
atoms are similar in various properties including, for example,
discrete anharmonic energy
levels~\cite{you2003quantum,rBlais04,rYou2011}, superconducting
atoms have some substantial advantages over natural atoms. These
include: (i) The spacing between energy levels, decoherence rates,
and coupling strengths between different circuit elements are
tunable by adjusting external parameters, thus, providing
flexibility for quantum information
processing~\cite{rBuluta11,rXiang13}; (ii) \new{The potential energies of natural
atoms have an inversion symmetry, leading to
a well-defined parity symmetry for each eigenstate. Thus,
an allowable dipole transition requires a parity change,
and can only occur between the two eigenstates having
different parities. However, the potential energies of superconducting artificial atoms can be controlled~\cite{rLiu05} by external parameters and, in turn, the inversion symmetry can
be broken or unbroken. When the symmetry is unbroken, each eigenstate of an artificial atom can have a well-defined parity symmetry and, thus, exhibit a transition
behavior similar to natural atoms. But when the symmetry is broken then there is no well-defined parity symmetry for each eigenstate and, thus, the microwave-induced
single-photon transition between any two eigenstates can
be possible.} That is, the selection rules can easily be
modified~\cite{rDeppe08}; (iii) The couplings between
different circuit elements can be strong, ultrastrong (for reviews
see~\cite{rYou2011,rXiang13}), or even deep
strong~\cite{Yoshihara17superconducting}, which, in particular, enable efficient
state preparation within a short time and with a high fidelity.

Specifically, based on circuit QED we realize a distant herald for
a multi-qubit Toffoli-like gate, whose the fidelity scaling is
quadratically improved as compared to unitary-dynamics-based gates
in cavities~\cite{Dgate1,cavitybasedgates}; moreover, heralded
controlled-Z gates even with more significant improvements can
also be achieved by using single-qubit operations.
\new{Note that the conditional measurement on the heralding atom is performed to remove the
detected errors from the quantum gate. Thus, these detected errors can only reduce the success probability but
not the gate fidelity. When the gate is successful, the
gate fidelity can still be very high as limited only by the
undetected errors.} In contrast to
previous works, a macroscopically distant herald for quantum gates
is proposed in superconducting circuits via the combined effect of
the dissipative and unitary dynamics, thereby having the potential
advantage of high efficiency in experimental scenarios. By
considering only the $d$ lowest energy levels of a superconducting
artificial atom, one can use them as a logical qudit for
performing quantum operations. In special cases, one can operate
with a qubit (for $d=2$), qutrit (for $d=3$), or quartit (ququart,
for $d=4$). Here we will analyze these special qudits.

This paper is organized as follows. In Sec.~\ref{se:model in
superconducting circuits}, we describe a physical model for
heralded quantum phase gates and propose a superconducting circuit
for its realization. In Sec.~\ref{se:effective master equation},
we derive the effective master equation which is used in
Sec.~\ref{se:Toffgates} to realize a multi-qubit Toffoli-like
gate. Working with single-qubit operations, Sec.~\ref{se:CZgates}
then presents a heralded controlled-Z gate. The last section is
our summary. A detailed derivation of the effective master
equation, which is applied in Sec.~III, is given in the Appendix.

\section{Physical model in superconducting circuits}
\label{se:model in superconducting circuits}
\begin{figure}[tbph]
\centering \includegraphics[width=8.0cm]{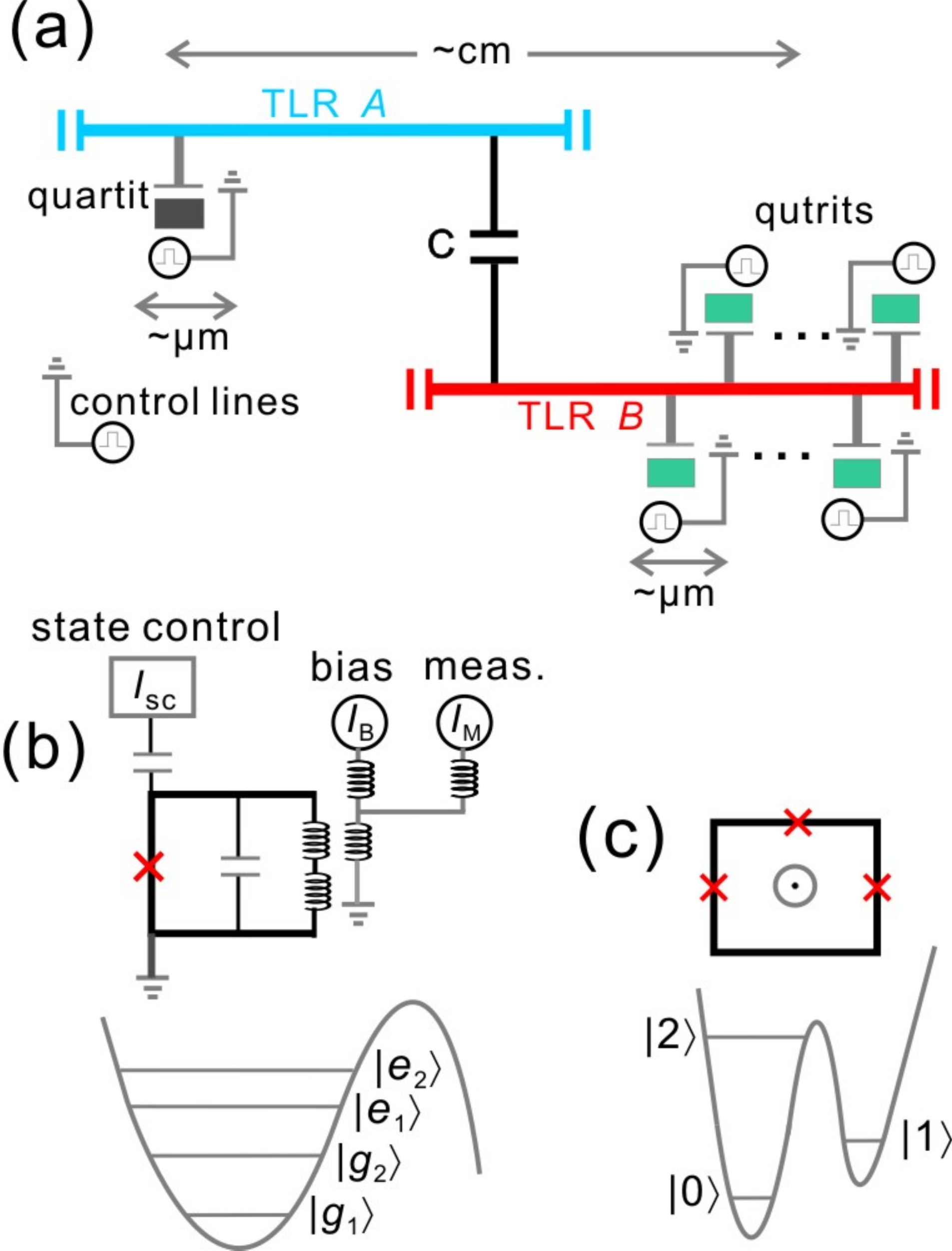} \caption{
(Color online) (a) Schematic diagram of a superconducting circuit
layout, which shows our implementation of a heralded
near-deterministic controlled multi-qubit Toffoli-like gate and,
in a special case, the two-qubit controlled-Z (CZ) gate. Two
transmission-line resonators (TLRs), labeled by $A$ and $B$, are
linearly coupled via a \new{capacitor}.
Resonator $A$ is coupled to a phase quartit (e.g., a four-level
qudit) via a capacitance, while resonator $B$ is coupled to $N$
identical flux qutrits (e.g., three-level qudits). Such circuit
elements can be controlled via ac and dc signals through the
control lines.  The distance between the qudit and qutrits can be
of order of cm due to the macroscopic length of the resonators.
Panel (b) shows an energy-level diagram for a prototypical phase
quartit, while panel (c) depicts that diagram for a typical flux
qutrit.  Here we assume that the energy potential for the quartit
is cubic and for the qutrit is an asymmetric double well. Note
that both flux and phase qudits can be tuned (for example, by
adjusting the flux bias in the qudit loop) to obtain exactly two,
three, or four levels. For example, following the method of
Refs.~\cite{martinis2002rabi,rNeeley09}, the phase qudit state can
be controlled by an ac signal $I_{\rm SC}$, while a dc signal
$I_{\rm B}$ is used to bias the circuit. Moreover, a short
measurement pulse $I_{\rm M}$ is applied to decrease the barrier
in the energy potential well of panel (b) to enable given upper
states to tunnel out of the well. These tunneling currents can be
detected by another SQUID which, for brevity, is not plotted here.
Thus, the proposed quantum gates can be realized, based solely on
flux or phase qudits, or other superconducting qudits. The
successful operation of this Toffoli-like gate is heralded by the
detection of the quartit in its ground state $|g_{1}\rangle$.}
\label{Figure1}
\end{figure}
\begin{figure*}[!ht]
\begin{center}
\includegraphics[width=17.0cm,angle=0]{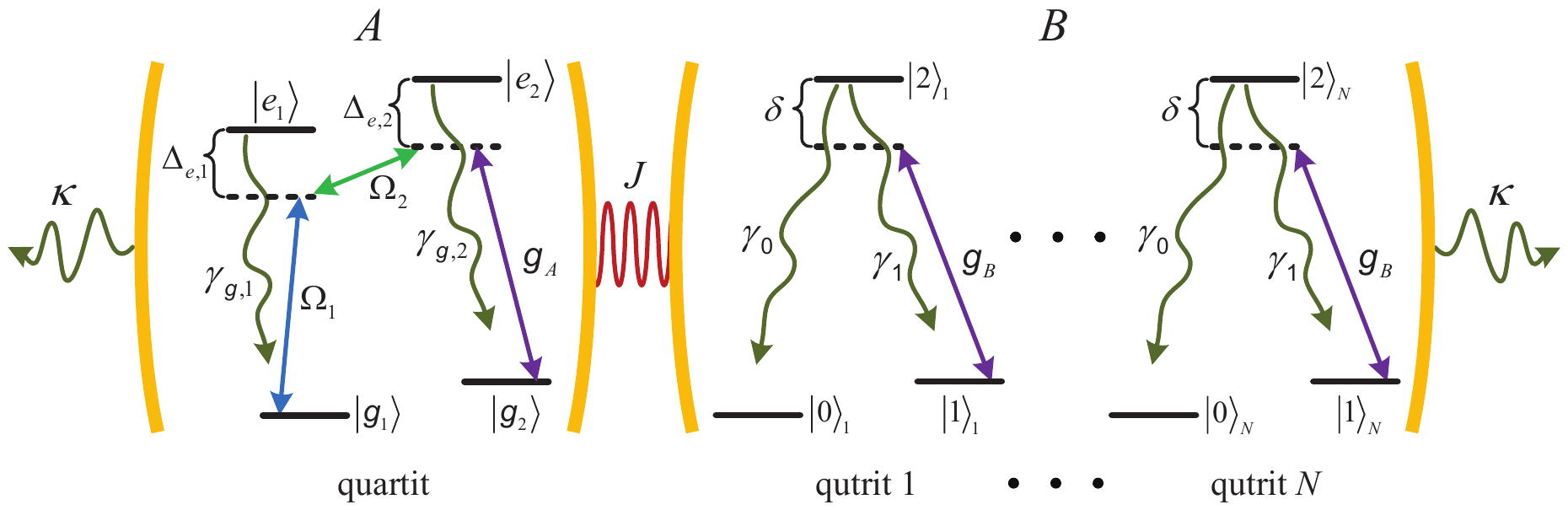}
\caption{(Color online) Energy-level diagram showing allowed
transitions and couplings in the circuit-QED system depicted in
Fig.~\ref{Figure1}. Two single-mode resonators, labeled by $A$ and
$B$ are coupled with a strength $J$. An auxiliary, quartit atom,
which acts as a heralding device, is confined in the resonator
$A$. Two microwave fields drive off-resonantly the transitions
$|g_{1}\rangle\rightarrow|e_{1}\rangle$ and
$|e_{1}\rangle\rightarrow|e_{2}\rangle$ of the auxiliary quartit,
with strengths $\Omega_{1}$ and $\Omega_{2}$, respectively.
Moreover, the states $|g_{2}\rangle$ and $|e_{2}\rangle$ are
coupled by the resonator mode $a_A$ with strength $g_{A}$. In the
resonator $B$, there are confined $N$ qutrit atoms, for which the
two lowest-energy levels can be treated as qubits. For each qutrit
atom, only the state $|1\rangle$ is coupled to $|2\rangle$ by the
resonator mode $a_B$ with strength $g_{B}$. Upon restricting our
discussion to sufficiently weak microwave drive $\Omega_{1}$, we
could adiabatically eliminate the excited states of the total
system to yield an effective Lindblad-type master equation,
which involves the ground states only. The resulting dynamics
allows for the realization of the controlled-phase gates, which
can successfully occur if the auxiliary quartit in the state
$|g_{1}\rangle$ is measured.}\label{Figure2}
\end{center}
\end{figure*}
\begin{table*}[!ht]
\caption{Basic notations used in this paper. Here, $x=1,2$ and
$z=0,1,2$.} \label{table}
\begin{tabular}{*{1}{p{3cm}<{\centering}}*{1}{p{12cm}<{\centering}}}
\hline \hline
notation & Meaning\\
\hline
$\omega_{g\left(e\right),x}$ & $|g_{x}\rangle$-($|e_{x}\rangle$-) level frequency  \\
$\omega_{z}$ & $|z\rangle$-level frequency   \\
$\omega_{c}$ & Common resonance frequency of resonators $A$ and $B$\\
$\omega_{m,x}$ & Microwave drive $x$ frequency \\
$\Omega_{x}$ & Microwave $x$ driving amplitude strength    \\
$g_{A}$ ($g_{B}$) & Coupling strength between the quartit (qutrit) atom and resonator $A$ ($B$)  \\
$J$ & Inter-resonator coupling strength   \\
$\gamma_{g,x}$ & Decay rate from level $|e_{x}\rangle$ to $|g_{x}\rangle$  \\
$\gamma_{x-1}$ & Decay rate from level $|2\rangle$ to $|x-1\rangle$    \\
$\gamma$ & Total decay rate, $\gamma=\gamma_{0}+\gamma_{1}$, of each qutrit atom \\
$\kappa$ & Resonator decay rate    \\
$C_{B}$ & Atom-resonator cooperativity, $C_{B}=g_{B}^{2}/\left(\kappa\gamma\right)$,    \\
$\Delta_{e,1}$ & Microwave detuning, $\Delta_{e,1}=\omega_{e,1}-\omega_{g,1}-\omega_{m,1}$ \\
$\Delta_{e,2}$ & Microwave detuning, $\Delta_{e,2}=\omega_{e,2}-\omega_{g,1}-\omega_{m,1}-\omega_{m,2}$ \\
$\delta$ & Resonator detuning, $\delta=\omega_{2}-\omega_1-\omega_{m,1}-\omega_{m,2}-\omega_{g,1}+\omega_{g,2}$   \\
\hline
\end{tabular}
\end{table*}


The basic idea underlying our protocol is schematically
illustrated in Figs.~\ref{Figure1} and~\ref{Figure2}.
Figure~\ref{Figure1} shows our proposal of the circuit-QED
implementation of the protocol based on superconducting qudits,
while the corresponding energy-level diagrams of the qudits are
depicted in Fig.~\ref{Figure2}.
Specifically, we consider two superconducting transmission-line
resonators $A$ and $B$, connected by a capacitor~\cite{Fitzpatrick2016observation}. The
coupling Hamiltonian, of strength $J$, for the two resonators can
be expressed as (hereafter we set $\hbar=1$)
\begin{equation}\label{eq1}
H_{AB}=J(a_{A}a_{B}^{\dag}+a_{A}^{\dag}a_{B}),
\end{equation}
where $a_{A}$ ($a_{B}$) is the annihilation operator of the
resonator $A$ ($B$).
We assume that superconducting artificial atoms, which are treated
as qudits (i.e., $d$-level quantum
systems)~\cite{rNeeley09,nori2009quantumfootball}, are coupled to
the resonators. Specifically, a superconducting phase quartit is
directly confined inside the resonator $A$, and is used as an
auxiliary quartit atom to herald the success of quantum gates.
Such a quartit consists of two ground levels $|g_{1}\rangle$ and
$|g_{2}\rangle$, as well as two excited levels $|e_{1}\rangle$ and
$|e_{2}\rangle$, depicted in Fig.~\ref{Figure1}(b). \new{Because the potential energy of the phase quartit is al-
ways broken, the quartit levels have no well-defined parity symmetry~\cite{Martinis09superconducting, rNeeley09,Liu14selection}. Thus, we can couple any two
levels by applied fields.} We assume that
the transition between $|g_{2}\rangle$ and $ |e_{2}\rangle$ is
coupled to the resonator $A$ by an inductance, with a Hamiltonian
\begin{equation}
H_{A}=g_{A}\left(a_{A}|e_{2}\rangle\langle
g_{2}|+\text{H.c.}\right), \label{eq2}
\end{equation}
where $g_{A}>0$ is a coupling strength.

In a similar manner, we couple the resonator $B$ to $N$
$\Lambda$-type qutrits, for example, superconducting flux
three-level atoms~\cite{rYu04,rLiu05}. Each qutrit consists of two
ground levels $|0\rangle$ and $|1\rangle$, together with one
excited level $|2\rangle$, depicted in Fig.~\ref{Figure2}(c). The
lowest two levels of an atomic qutrit are treated as qubit states.
\new{With current technologies, superconducting atoms can be made almost identical. Thus, for simplicity, we can assume that these qutrits
are identical and have the same coupling, of strength $g_B$,
to the resonator $B$.} Such a coupling can be
ensured by adjusting the control signals on the qutrits and by
tuning the separation between any two qutrits to be much smaller
than the wavelength of the resonator $B$, respectively. The
corresponding Hamiltonian is
\begin{equation}\label{eq3}
H_{B}=g_{B}\sum\limits_{k=1}^{N}\left(a_{B}|2\rangle_{k} \langle
1|+\text{H.c.}\right).
\end{equation}
where $k$ labels the qutrits and $g_{B}>0$ is the coupling
strength of the resonator $B$. Note that the direct dipole-dipole
coupling between the qutrits has been neglected, owing to their
large spatial separations. Nevertheless, these qutrits can
interact indirectly via the common field $a_{B}$ of the resonator
$B$, analogously to the model of three-level quantum dots in a
resonator studied for quantum-information processing in
Refs.~\cite{Imamoglu99,Miranowicz02}. Because the Josephson
junctions are nonlinear circuits elements~\cite{rYou2011}, and
therefore the resulting levels are highly anharmonic compared to
the driving strengths as well as to the atom-resonator coupling
strengths, all transitions of the quartit and the qutrits can be
driven or coupled separately by the control lines, as shown in
Fig.~\ref{Figure2}(a).

We also assume that a microwave field at the frequency
$\omega_{m,1}$ drives the
$|g_{1}\rangle\leftrightarrow|e_{1}\rangle$ transition, with a
driving strength $\Omega_{1}$ and at the same time, the excited
states $|e_{1}\rangle$ and $|e_{2}\rangle$ are also coupled by
means of a microwave field at the frequency $\omega_{m,2}$, with a
coupling strength $\Omega_{2}$. The interaction Hamiltonian
describing the effect of these external drives reads as
\begin{equation}
H_{D}=\frac{1}{2}\left(\Omega_{1}e^{i\omega_{m,1}t}|g_{1}\rangle\langle
e_{1}|+\Omega_{2}e^{i\omega_{m,2}t}|e_{1}\rangle\langle
e_{2}|+\text{H.c.}\right).\label{H_D}
\end{equation}
Define $\omega_{g,x}$, $\omega_{e,x}$, and $\omega_{z}$ as the
frequencies of the atomic levels $|g_{x}\rangle$, $|e_{x}\rangle$,
and $|z\rangle$, respectively, with $x=1,2$ and $z=0,1,2$. Thus,
the total Hamiltonian for our system is
\begin{equation}
H_{T}=H_{0}+H_{A}+H_{B}+H_{AB}+H_{D},
 \label{H_total}
\end{equation}
where
\begin{align}\label{eq:freeH}
H_{0}=&\sum_{x=1,2}\left(\omega_{g,x}|g_{x}\rangle\langle g_{x}|+\omega_{e,x}|e_{x}\rangle\langle e_{x}|\right) \nonumber\\
&+\sum_{k=1}^{N}\sum_{z=0,1,2}\omega_{z}|z\rangle_{k}\langle
z|+\omega_{c}\left(a_{A}^{\dag}a_{A}+a_{B}^{\dag}a_{B}\right),
\end{align}
is the free Hamiltonian.

Upon introducing the symmetric and antisymmetric optical modes,
$a_{\pm}=\left(a_{A}\pm a_{B}\right)/\sqrt{2}$, the total
Hamiltonian in a proper rotating frame reads
$H_{T}=H_{e}+V+V^{\dag}$, with
\begin{align}\label{eq:H}
H_{e}=&\sum_{k=1}^{N}\left\{\delta|2\rangle_{k}\langle 2|+\frac{g_{B}}{\sqrt{2}}\left[\left(a_{+}-a_{-}\right)|2\rangle_{k}\langle1|+\text{H.c.}\right]\right\}\nonumber\\
&+\Delta_{e,1}|e_{1}\rangle\langle e_{1}|+\Delta_{e,2}|e_{2}\rangle\langle e_{2}|+2Ja_{+}^{\dag}a_{+}\nonumber\\
&+\frac{g_{A}}{\sqrt{2}}\left[\left(a_{+}+a_{-}\right)|e_{2}\rangle\langle g_{2}|+\text{H.c.}\right]\nonumber\\
&+\frac{\Omega_{2}}{2}\left(|e_{2}\rangle\langle
e_{1}|+\text{H.c.}\right),
\end{align}
and
\new{\begin{equation}
V=\frac{\Omega_{1}}{2}|e_{1}\rangle\langle g_{1}|.
\end{equation}}
Note that we have applied the rotating-wave approximation (RWA), which directly
drops the fast oscillating terms of the total Hamiltonian. The
detunings are defined as (see Fig.~\ref{Figure2}):
\begin{equation}
\delta=\omega_{2}-\omega_1-\omega_{m,1}-\omega_{m,2}-\omega_{g,1}+\omega_{g,2},
\end{equation}
\begin{equation}
\Delta_{e,1}=\omega_{e,1}-\omega_{g,1}-\omega_{m,1},
\end{equation}
\begin{equation}
\Delta_{e,2}=\omega_{e,2}-\omega_{g,1}-\omega_{m,1}-\omega_{m,2},
\end{equation}
In Eq.~(\ref{eq:H}) we have assumed
\begin{equation}
\omega_{c}=\omega_{2}-\delta-\omega_{1}+J,
\end{equation}
such that the three-photon Raman transition between the levels
$|g_{1}\rangle\leftrightarrow|e_{1}\rangle
\leftrightarrow|e_{2}\rangle\leftrightarrow|g_{2}\rangle$, is
resonant when mediated by the antisymmetric mode $a_{-}$, but is
detuned by $2J$ when mediated by the symmetric mode $a_+$. We
further assume that $|e_{1}\rangle$ and $|e_{2}\rangle $ decay to
$|g_{1}\rangle$ and $|g_{2}\rangle$, respectively, with rates
$\gamma_{g,1}$ and $\gamma_{g,2}$, and for each qutrit atom,
$|2\rangle$ can decay either to $|0\rangle$ with a rate
$\gamma_{0}$ or to $|1\rangle$ with a rate $\gamma_1$. In
addition, both resonators are assumed to have the same decay rate
$\kappa$. All basic symbols used in this paper are shown in
Table~\ref{table}.

\section{Master equation}
\label{se:effective master equation}

Here we present a standard approach based on the master
equation in the Lindblad form to study the dissipative dynamics of
our system. The master equation is valid under a few fundamental
assumptions which
include~\cite{CarmichaelBook,ScullyBook,BreuerBook}: (A1) the
approximation of the weak coupling between the analyzed system and
its reservoir (environment), (A2) the Markov approximation, and
(A3) the secular approximation. The approximations (A1) and (A2)
are often referred to as the Born-Markov approximation. By
applying (A2), one assumes that the environmental-memory effects
are short-lived, such that the system evolution depends only on
its present state. This approximation is valid if the
environmental correlation time (which can be evaluated by the
decay timescale of the two-time correlation functions of the
environmental operators coupled to the system) is much shorter
than a typical system-evolution timescale over which the system
experiences a significant evolution. For example, the Born-Markov
approximation is valid if an environment is large and weakly
coupled to a system. The approximation (A3) is applied to cast a
given Markovian master equation into the Lindblad form. This
corresponds to ignoring fast-oscillating terms in the master
equation based on (A1) and (A2). Thus, (A3) is sometimes called
the RWA, although it is usually applied at the level of a given
master equation, and not necessarily at the level of the
system-reservoir interaction Hamiltonian. There are numerous
references showing the excellent agreement between the
experimental and theoretical results based upon the master
equation in the Lindblad form describing the lossy interaction of
quantum systems (including superconducting artificial atoms) and
resonator fields (see, e.g.,~\cite{Clerk10introduction,Gu2017microwave} and references
therein). The validity of these approximations for a single-qudit
version of our system was also experimentally analyzed in
Ref.~\cite{Dentanglement4}.

The standard master equation in the Lindblad form for the system
described by the Hamiltonian given in Eq.~(\ref{H_total}),
assuming the zero-temperature environment (bath), can be given
by~\new{\cite{CarmichaelBook, ScullyBook,BreuerBook}}:
\begin{eqnarray}
\label{eq:fullmasequa}
\dot{\rho}_{T}\left(t\right)&=&i\left[\rho_{T}\left(t\right),H_{T}\right]
+\frac{1}{2}\sum_{j}\Big[2L_{j}\rho_{T}\left(t\right)L_{j}^{\dag} \nonumber \\
&&
-\rho_{T}\left(t\right)L_{j}^{\dag}L_{j}-L_{j}^{\dag}L_{j}\rho_{T}\left(t\right)\Big],
\label{ME}
\end{eqnarray}
where $\rho_{T}\left(t\right)$ is the density operator of the
total system. The Lindblad operators associated with the resonator
decay and atomic spontaneous emission are accordingly given by
\begin{align}\label{eq:lindbladoperators}
&L_{\pm}=\sqrt{\kappa}a_{\pm}, \nonumber\\
&L_{g,x}=\sqrt{\gamma_{g,x}}|g_{x}\rangle\langle e_{x}|, \\
&L_{k,x-1}=\sqrt{\gamma_{x-1}}|x-1\rangle\;_{k}\langle 2|,
\nonumber
\end{align}
where $k$ labels the qutrit atoms, and $x=1,2$.

We now consider a weak microwave drive $\Omega_{1}$, so that \new{$\left\{\Omega_{1}/\Delta_{e,1},\Omega_{1}/g_{A\left(B\right)}\right\}\ll1$. In this situation,} both
the resonator modes and excited atomic states can be adiabatically
eliminated if the system is initially in its ground state.
Following the procedure in Ref.~\cite{effmaseq}, the dynamics is
therefore described by the effective Hamiltonian,
\begin{equation}\label{eq:formalism-for-effHam}
H_{\text{eff}}=-\frac{1}{2}V^{\dag}\left[H_{\text{NH}}^{-1}+\left(H_{\text{NH}}^{-1}\right)^{\dag}\right]V,
\end{equation}
and the effective Lindblad operators,
\begin{equation}\label{eq:formalism-for-effLind}
L_{\text{eff}}^{j}=L_{j}H_{\text{NH}}^{-1}V.
\end{equation}
Here,
\begin{equation}\label{eq:no-jump-Ham}
H_{\text{NH}}=H_{e}-\frac{i}{2}\sum_{j}L_{j}^{\dag}L_{j}
\end{equation}
accounts for the no-jump Hamiltonian, where the sum runs over all
dissipative processes as mentioned in
Eq.~(\ref{eq:lindbladoperators}). The effective Lindblad master
equation then has the form
\begin{align}\label{eq:effmastereq}
\dot{\rho}\left(t\right)=\;&i\left[\rho\left(t\right),H_{\text{eff}}\right]+\frac{1}{2}\sum_{j}\bigg\{2L^{j}_{\text{eff}}\,\rho\left(t\right) \left(L_{\text{eff}}^{j}\right)^{\dag}   \nonumber\\
&-\left[\left(L_{\text{eff}}^{j}\right)^{\dag}L_{\text{eff}}^{j}\,\rho\left(t\right)
+\rho\left(t\right)\left(L_{\text{eff}}^{j}\right)^{\dag}L_{\text{eff}}^{j}\right]\bigg\},
\end{align}
assuming the reservoir at zero temperature. Here
$\rho\left(t\right)$ is the density operator of the quartit and
qutrit atoms.

As explained in detail in the Appendix, when working within the
limits $\Omega_{2}\ll\Delta_{e,1}$ and $\kappa\ll J$, we can more
explicitly obtain the effective Hamiltonian
\begin{equation}\label{eq:effH}
H_{\text{eff}}=|g_{1}\rangle\langle
g_{1}|\otimes\sum_{n=0}^{N}\Delta_{n}\mathcal{P}_{n},
\end{equation}
where $\mathcal{P}_{n}$ represents a projector onto a subspace
characterized by $n$ atomic qutrits in $|1\rangle$, and
\begin{equation}\label{eq:deltan_a}
\Delta_{n}=-\frac{\widetilde{\Omega}^{2}}{4\gamma}\text{Re}\left\{\frac{1}{\mathcal{Z}_{n}}\left(i\widetilde{\delta}+nC_{B}\right)\right\}
\end{equation}
\new{refers to an AC stark shift} with
\begin{equation}
\mathcal{Z}_{n}=i\widetilde{\delta}\widetilde{\Delta}_{e,2}+C_{B}\left(\alpha\widetilde{\delta}
+n\widetilde{\Delta}_{e,2}\right)-n\alpha C_{B}^{2}/G.
\end{equation}
Here, we have defined the overall decay rate
$\gamma=\gamma_{0}+\gamma_{1}$, of each qutrit atom, the effective drive
$\widetilde{\Omega}=\Omega_{1}\Omega_{2}/\left(2\Delta_{e,1}\right)$,
\new{and the following dimensionless quantities: the atom-resonator cooperativity
$C_{B}=g^{2}_{B}/\left(\kappa\gamma\right)$, the strength $G=J/\kappa$, the complex detunings $\widetilde{\delta}=\delta/\gamma-i/2$,
$\widetilde{\Delta}_{e,2}=\Delta_{e,2}/\gamma-i\beta/2$, and the parameters}
\begin{equation}
\alpha=\left(g_{A}/g_{B}\right)^2, \quad
\beta=\gamma_{g,2}/\gamma.
\end{equation}
In all our numerical simulations we set $\alpha=\beta=1$. The
term, $-\Omega^2_{1}/\left(4\Delta_{e,1}\right)$, of $\Delta_{n}$
has been removed since it only causes an overall phase.
Equation~(\ref{eq:effH}) indicates that the time evolution under
the effective Hamiltonian gives rise to a dynamical phase
imprinted onto each state of the qutrits, while making the state
of the quartit atom unchanged. Correspondingly, the effective
Lindblad operators are found to be
\begin{align}
L_{\text{eff}}^{\pm}&=|g_{2}\rangle\langle g_{1}|\otimes\sum_{n=0}^{N}r_{\pm,n}\mathcal{P}_{n},\label{eq:Lpm}\\
L_{\text{eff}}^{g,x}&=|g_{x}\rangle\langle g_{1}|\otimes\sum_{n=0}^{N}r_{g_{x},n}\mathcal{P}_{n},\label{eq:Lgx}\\
L_{\text{eff}}^{k,x-1}&=|g_{2}\rangle\langle
g_{1}|\otimes\sum_{n=1}^{N}r_{x-1,n}|x-1\rangle_{k}\langle1|\mathcal{P}_{n},\label{eq:Lkx}
\end{align}
with $x=1,2$. Here the effective decay rates $r_{\pm,n}$,
$r_{g_{x},n}$, and $r_{x-1,n}$ are expressed, respectively, as
\begin{align}\label{eq:effectivelindbladoperators_a}
&r_{+,n}=\frac{\widetilde{\Omega}\sqrt{\alpha C_{B}}}{4G\mathcal{Z}_{n}\sqrt{2\gamma}}\left(i\widetilde{\delta}+2nC_{B}\right),\nonumber\\
&r_{-,n}=-\frac{\widetilde{\Omega}\sqrt{\alpha C_{B}}}{\mathcal{Z}_{n}\sqrt{2\gamma}}\left[i\widetilde{\delta}-nC_{B}/\left(2G\right)\right],\nonumber\\
&r_{g_{1},n}=\frac{\Omega_{1}\sqrt{\gamma_{g,1}}}{2\Delta_{e,1}},\\
&r_{g_{2},n}=-\frac{\widetilde{\Omega}\sqrt{\beta}}{2\mathcal{Z}_{n}\sqrt{\gamma}}\left(i\widetilde{\delta}+nC_{B}\right), \nonumber\\
&r_{x-1,n}=-\frac{\widetilde{\Omega}\sqrt{\alpha\gamma_{x-1}}}{2\gamma\mathcal{Z}_{n}}C_{B},
\nonumber
\end{align}
as derived in the Appendix. According to these effective Lindblad
operators, we find that all dissipative processes, except the one
corresponding to $L_{\text{eff}}^{g,1}$ independent of $n$, cause
the $|g_{1}\rangle\rightarrow|g_{2}\rangle$ decay. The errors
induced by this decay are detectable by measuring the state of the
quartit atom, and can be removed by conditioning on the quartit
atom being measured in $|g_{1}\rangle$. Upon solving the
effective master equation of Eq.~(\ref{eq:effmastereq}), the
probability of detecting the quartit atom in the state
$|g_{1}\rangle$ is given by
\begin{equation}\label{eq:probps}
P=\sum_{n=0}^{N}\text{Tr}\left[\left(|g_{1}\rangle\langle
g_{1}|\otimes \mathcal{P}_{n}\right)\rho\left(t\right)\right],
\end{equation}
where $\text{Tr}$ is the trace operation over the subspace spanned by the
ground states of the quartit and qutrit atoms, and
$\sum_{n=0}^{N}\mathcal{P}_{n}=\mathcal{I}_{G}$ is the identity operator
acting on the ground state manifold of qutrit atoms. Such a
detection could be performed using the method developed in
Refs.~\cite{martinis2002rabi,rNeeley09} for a phase qudit, which
we have briefly described in the caption of Fig.~\ref{Figure1}. We
note that also other schemes for the detection of phase and flux
qudit states can be applied including the dispersive read-out
methods~\cite{wallraff05approaching,
mallet2009single,chen2012multiplexed}. After this measurement, the
conditional density operator of the qutrits is then
\begin{multline}\label{eq:qubitdmatrix}
\rho_{\text{qutrit}}\left(t\right)=\frac{1}{P}\sum_{n,n'=0}^{N}e^{-i\left(\Delta_{n}-\Delta_{n'}\right)t}e^{-\left(\Gamma_{n}+\Gamma_{n'}\right)t/2}      \\
\times \mathcal{P}_{n}\left[\langle
g_{1}|\rho\left(0\right)|g_{1}\rangle\right]\mathcal{P}_{n'},
\end{multline}
with the following total decay rate:
\begin{multline}\label{eq:total-decay-rate}
\Gamma_{n}=|r_{+,n}|^{2}+|r_{-,n}|^{2}+|r_{g2,n}|^{2}+n\left(|r_{0,n}|^{2}+|r_{1,n}|^{2}\right).
\end{multline}
Thus, by measuring the quartit atom and referring to the $P$ as
the success probability, we could realize heralded quantum
controlled phase gates as discussed below. In this approach,
the detectable errors only reduce the success probability, instead
of reducing the gate fidelity. If we successfully measure the
quartit in $|g_{1}\rangle$, then the resulting gate has a very
high fidelity, which is limited only by the undetectable errors
induced, for example, by the differences between the decay rates
$\Gamma_{n}$ for different states of the qutrit atoms. Thus, the
underlying key idea is to remove the detectable errors by a
heralding measurement and, then, to achieve quantum gates with
very high fidelities in two macroscopically-distant resonators.

\begin{table}[ht]
\caption{The action of a heralded multi-qubit Toffoli-like
gate. Note that
$\Delta_{0}\neq\Delta_{1}=\Delta_{2}=\cdots=\Delta_{N}$, and we
have also ignored an overall phase, $\exp(-i\pi)$, of the output
states.} \label{table2}
\begin{tabular}{*{1}{p{1cm}<{\centering}}*{1}{p{2.3cm}<{\centering}}*{1}{p{2.4cm}<{\centering}}*{1}{p{2.3cm}<{\centering}}}
\hline \hline $n$ & Input states & $\xrightarrow{\text{Time
evolution}}$ & Output states\\ [2pt]\hline $0$ &
$|000\cdots0\rangle$ &
$\xrightarrow{\exp{\left(-i\Delta_{0}t_{\text{Toff}}\right)}}$ &
$-|000\cdots0\rangle$ \\ [2pt] \hline \multirow{5}{*}{1} &
$|100\cdots0\rangle$ &
\multirow{5}{*}{$\xrightarrow{\exp{\left(-i\Delta_{1}t_{\text{Toff}}\right)}}$}
& $|100\cdots0\rangle$ \\ [2pt] & $|010\cdots0\rangle$ &  &
$|010\cdots0\rangle$ \\ [2pt] & $\vdots$ &  & $\vdots$ \\ [2pt] &
$|0\cdots001\rangle$ &  & $|0\cdots001\rangle$ \\ [2pt]\hline
\multirow{5}{*}{2} & $|110\cdots0\rangle$ & \multirow{5}{*}{$\xrightarrow{\exp{\left(-i\Delta_{2}t_{\text{Toff}}\right)}}$} & $|110\cdots0\rangle$ \\
& $|101\cdots0\rangle$ &  & $|101\cdots0\rangle$ \\ [2pt] &
$\vdots$ &  & $\vdots$ \\ [2pt] & $|0\cdots011\rangle$ &  &
$|0\cdots011\rangle$ \\ [2pt]\hline $\vdots$ & $\vdots$ & $\vdots$
& $\vdots$ \\ [2pt] \hline $N$ & $|111\cdots1\rangle$ &
$\xrightarrow{\exp{\left(-i\Delta_{N}t_{\text{Toff}}\right)}}$ &
$|111\cdots1\rangle$ \\ [2pt]\hline
\end{tabular}
\end{table}
\begin{table}[ht] \caption{The action of the heralded
controlled-Z gate. Note that $\Delta_{0}$, $\Delta_{1}$, and
$\Delta_{2}$ are not equal to each other. Thus, some unitary
operations should be applied on qubits to achieve their proper
phase evolution.} \label{table3}
\begin{tabular}{*{1}{p{1cm}<{\centering}}*{1}{p{2.3cm}<{\centering}}*{1}{p{2.4cm}<{\centering}}*{1}{p{2.3cm}<{\centering}}}
\hline \hline
$n$ & Input states & $\xrightarrow{\text{Time evolution}}$ & Output states\\ [2pt]
\hline
$0$ & $|00\rangle$ & $\xrightarrow{\exp{\left(-i\Delta_{0}t_{\text{CZ}}\right)}}$ & $|00\rangle$ \\ [2pt]
\hline
\multirow{2}{*}{1} & $|10\rangle$ & \multirow{2}{*}{$\xrightarrow{\exp{\left(-i\Delta_{1}t_{\text{CZ}}\right)}}$} & $|10\rangle$ \\ [2pt]
& $|01\rangle$ &  & $|01\rangle$ \\ [2pt]
\hline
2 & $|11\rangle$ & $\xrightarrow{\exp{\left(-i\Delta_{2}t_{\text{CZ}}\right)}}$ & $-|11\rangle$ \\ [2pt]
\hline
\end{tabular}
\end{table}
\section{Heralded multi-qubit\\ Toffoli-like gate}
\label{se:Toffgates} In this section we will demonstrate a
heralded near-deterministic multi-qubit Toffoli-like gate, which
is defined as the multi-qubit controlled-Z gate. Thus, the action
of our Toffoli-like gate on $N$ qubits in a state
$\ket{\psi}=\ket{q_1,q_2,...,q_N}$ is given by
$\ket{\psi}\rightarrow(-1)^{\left(q_1\oplus1\right)\left(q_2\oplus1\right)\cdots
\left(q_N\oplus1\right)} \ket{\psi}$. In our case these logical
qubits correspond to the two lowest states of the qutrits in the
resonator $B$. The successful action of this Toffoli-like gate is
heralded by the detection of the quartit in its ground state
$|g_{1}\rangle$. Note that the standard three-qubit Toffoli gate
is defined as the double-controlled NOT (CCNOT) gate and given by
the map~\cite{chuang}: $\ket{q_1,q_2,q_3}\rightarrow
\ket{q_1,q_2,q_3\oplus q_1q_2}$, rather than the map
$\ket{q_1,q_2,q_3}\rightarrow
(-1)^{\left(q_1\oplus1\right)\left(q_2\oplus1\right)\left(q_3\oplus1\right)}
\ket{q_1,q_2,q_3}$, which is applied here as in, e.g.,
Ref.~\cite{Dgate1}.

In order to realize our multi-qubit Toffoli-like gate, we can make
$\Delta_{n>0}$ independent of $n$, and ensure that
$|\Delta_{0}|\ll|\Delta_{n>0}|$. To this end, we tune $\delta=0$
and $\Delta_{e,2}/\gamma=\alpha C_{B}\left(R+1/G\right)$, where
\begin{equation}
R=\sqrt{\frac12\left(\frac1{G^2}+\frac{\beta}{\alpha
C_{B}}+\frac1{C_{B}}\right)}.
\end{equation}
In the limit of $\min\{G,C_{B}\}\gg1$, this choice can lead
to
\begin{eqnarray}
\Delta_{0}&=&-\frac{\widetilde{\Omega}^2}{4\gamma}\frac{1}{\alpha
C_{B}}\left(R+1/G\right),
\\
\Delta_{n>0}&=&-\frac{\widetilde{\Omega}^2}{4\gamma}\frac{1}{\alpha
C_{B}R},
\end{eqnarray}
which satisfies the condition $|\Delta_{0}|\ll|\Delta_{n>0}|$.
Therefore, our $N$-qubit Toffoli-like gate can be achieved by
applying a driving pulse of duration
$t_{\text{Toff}}=\pi/|\Delta_{n>0}|$. Up to an overall phase, such
a gate flips the phase of the qubit state with all qubits in
$|0\rangle$, but has no effect on the other qubit states. However,
the particular detunings $\delta$ and $\Delta_{e,2}$ also yield
\begin{eqnarray}\label{eq:ToffGamma}
\Gamma&\equiv&\Gamma_{0}=\Gamma_{1}=\frac{\widetilde{\Omega}^2}{2\gamma}\frac{1}{\alpha
C_{B}},
\\
\Gamma_{n>1}&=&\frac{\widetilde{\Omega}^2}{4\gamma}\frac{1}{\alpha
C_{B}}\left(2-\frac{1-1/n}{C_{B}R^2}\right),
\end{eqnarray}
again in the limit of $\min\{G,C_{B}\}\gg1$. According to Eq.
(27), the decay factor
$\exp[-\left(\Gamma_{n}+\Gamma_{n'}\right)/2]$  cannot be completely
removed conditional on the quartit being measured in
$|g_{1}\rangle$, thus, leading to gate errors.

\begin{figure}[tbph]
\begin{center}
\includegraphics[width=8.2cm,angle=0]{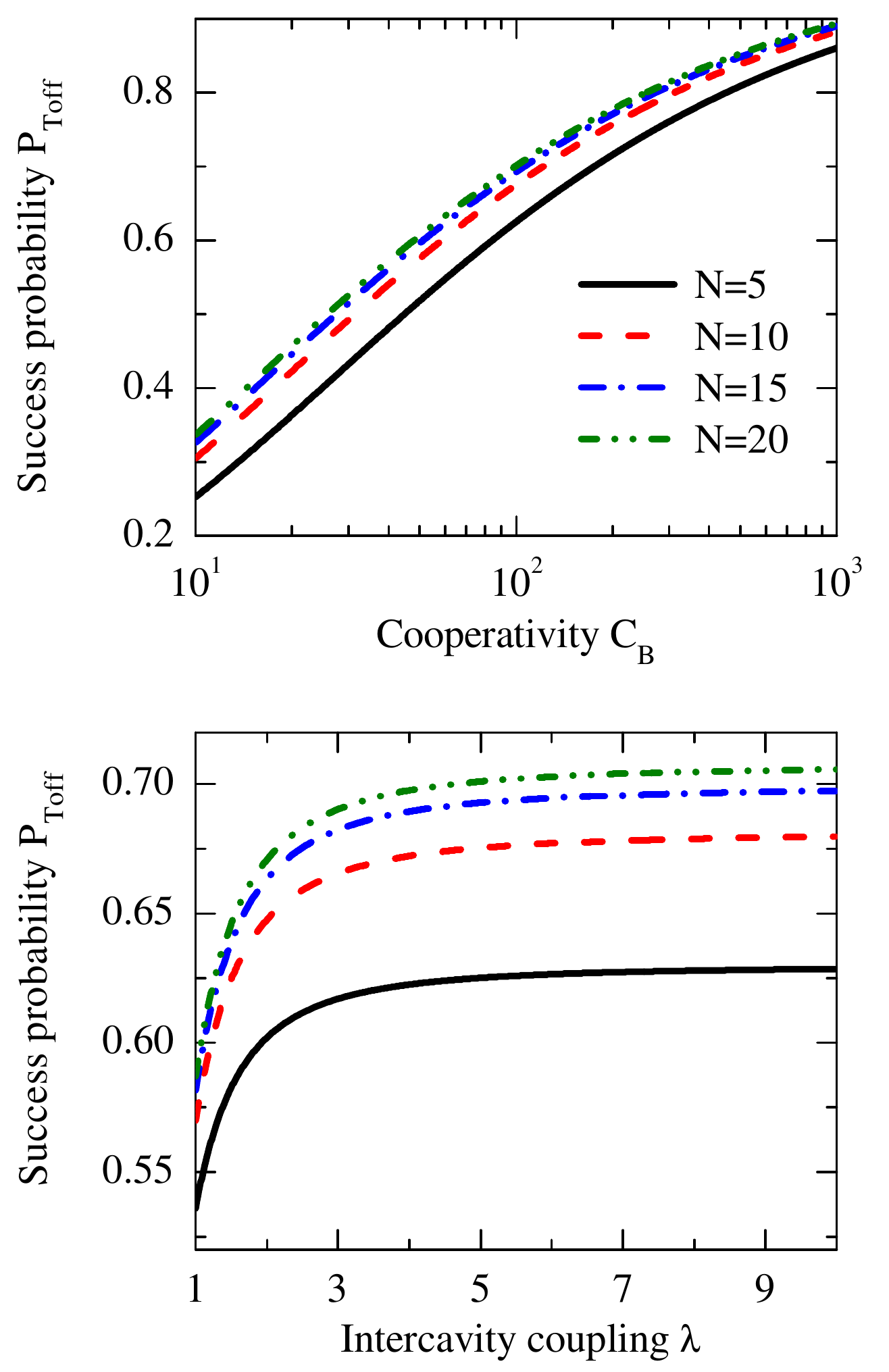}
\caption{(Color online) Success probability, $P_{\text{Toff}}$, of
our multi-qubit Toffoli-like gate as a function of (a) the
cooperativity $C_{B}$ or (b) the inter-resonator-coupling strength
$\lambda$, and for $N=5$ (solid black curves), $10$ (dashed red
curves), $15$ (dashed-dotted blue curves), and $20$ (dashed-double
dotted olive curves) three-level atoms. Here, we have assumed that
$\lambda=5$ in (a) and $C_{B}=100$ in (b). In both panels, the
damping rates are set as \new{$\gamma_{g,1}=\gamma_{g,2}=\gamma$},
$\gamma_{0}=\gamma_{1}=0.5\gamma$, \new{$g_{A}=g_{B}$, $C_{B}=g_{B}^{2}/\left(\kappa\gamma\right)$, $\lambda=J/\left(\kappa\sqrt{C_{B}}\right)$,} and
$\alpha=\beta=1$.}\label{Figure3}
\end{center}
\end{figure}

In order to quantify the quality of this gate, we need to define a
conditional fidelity as
\begin{equation}
F_{\text{Toff}}=\langle
\phi|\rho_{\text{qubit}}\left(t_{\text{Toff}}\right)|\phi\rangle,
\end{equation}
where $|\phi\rangle$ is assumed to be the desired state after the
gate operation. Correspondingly, the gate error is
characterized by $1-F_{\text{Toff}}$.  Considering a generic
initial state of the quartit and qutrit atoms,
\begin{equation}\label{eq:atomicinitialstate}
|\Phi\rangle_{\text{ini}}=|g_{1}\rangle\left[\frac{1}{2^{N/2}}\prod_{k=1}^{N}\left(|0\rangle+|1\rangle\right)_{k}\right],
\end{equation}
the success probability and the conditional fidelity is given by
\begin{align}
P_{\text{Toff}}&=\frac{1}{2^{N}}\sum_{n=0}^{N}C_{N}^{n}\exp\left(-\Gamma_{n}t_{\text{Toff}}\right),\\
F_{\text{Toff}}&=\frac{1}{2^{2N}P_{\text{Toff}}}\left[\sum_{n=0}^{N}C_{N}^{n}\exp\left(-\Gamma_{n}t_{\text{Toff}}/2\right)\right]^{2},
\end{align}
respectively, with $C_{N}^{n}=N!/\left[n!\left(N-n\right)!\right]$
being the binomial coefficient. Again in the limit
$\left\{G,C_{B}\right\}\gg1$, we have
$\Gamma_{n}t_{\text{T}}\ll0$, which in turn results in
\begin{align}
\label{eq:Toffsuccessprob}
P_{\text{Toff}}&=1-T_{p}\frac{\pi}{\sqrt{C_{B}}},\\
\label{eq:Tofferror}
F_{\text{Toff}}&=1-T_{f}\frac{\pi^{2}}{C_{B}},
\end{align}
where the scaling factors $T_{p}$ and $T_{f}$ are written as
\begin{equation}
T_{p}=2r+\frac{1}{r}\left[\frac{1}{2^N}\left(1+S_{1}\right)-1\right],
\end{equation}
\begin{equation}
T_{f}=\frac{1}{2^{N+2}r^{2}}\left[\left(1+S_{2}\right)-\frac{1}{2^N}\left(1+S_{1}\right)^2\right],
\end{equation}
respectively. Here,
$r=\sqrt{\left(1/\lambda^{2}+\beta/\alpha+1\right)/2}$ with
$\lambda=G/\sqrt{C_{B}}$, and
$S_{x}=\sum_{n=1}^{N}C_{N}^{n}/n^{x}$ for $x=1,2$. Together with a
success probability
\begin{figure}[!ht]
\begin{center}
\includegraphics[width=8.2cm,angle=0]{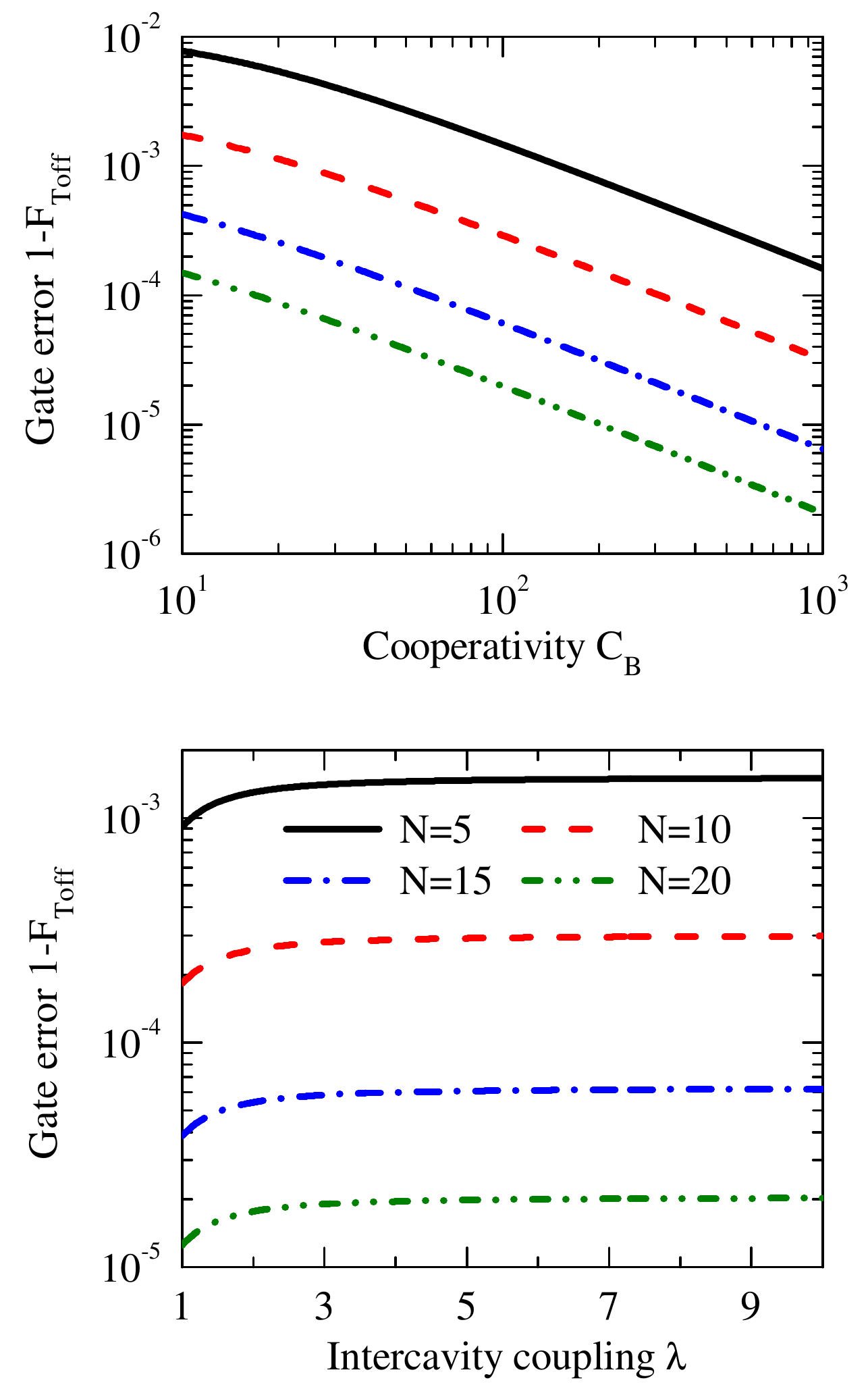}
\caption{(Color online) Analytical results for the gate error
(gate infidelity), $1-F_{\text{Toff}}$, of our multi-qubit
Toffoli-like gate as a function of (a) the cooperativity $C_{B}$
or (b) the inter-resonator coupling strength $\lambda$, and for
$N=5$ (solid black curves), $10$ (dashed red curves), $15$ (
dashed-dotted blue curves), and $20$ (dashed-double dotted olive
curves) three-level atoms (qutrits). The other parameters are also
set to be the same as in Fig.~\ref{Figure3}. There the success
probability is in a one-to-one correspondence with the gate error
here.}\label{Figure4}
\end{center}
\end{figure}
\begin{equation}
P_{\text{Toff}}\propto1-1/\sqrt{C_{B}}.
\end{equation}
Thus, the proposed protocol for our multi-qubit Toffoli-like gate
has an error scaling as
\begin{equation}\label{eq:Tofferrorscaling}
1-F_{\text{Toff}}\propto 1/C_{B},
\end{equation}
which is a quadratic improvement as compared to gate protocols
making use of coherent unitary dynamics in
cavities~\cite{cavitybasedgates}, as explained in
Ref.~\cite{Dgate1}. The latter gate suffers errors from the
resonator decay and atomic spontaneous emission. Instead our
protocol exploits the combined effect of the unitary and
dissipative processes, thus, resulting in the above improvement.
In fact, the atom-resonator cooperativity $C_{B}$ could
experimentally reach $>10^{4}$ in superconducting
circuits~\cite{rXiang13,niemczyk2010circuit}, thus, making the
gate error very close to zero and the success probability close to
unity.

The success probability is plotted as a function of either the
cooperativity $C_{B}$ or the inter-resonator coupling $J$,
illustrated in Fig.~\ref{Figure3}. There we have assumed that
$\gamma_{g,1}=\gamma_{g,2}$, $\gamma_{0}=\gamma_{1}=0.5\gamma$,
and the quartit atom is the same as the qutrit atoms, such that
$\alpha=\beta=1$. Similarly, we also plot the corresponding gate
error in Fig.~\ref{Figure4}. As expected, we find that increasing
the cooperativity not only makes the success probability very high
[see Fig.~\ref{Figure3}(a)], but it also makes the gate error very
low [see Fig.~\ref{Figure4}(a)]. For example, the success
probability of up to $\sim0.9$ and the gate error of up to
$\sim2.0\times 10^{-6}$ can be achieved when $N=20$, $\lambda=5$
and $C_{B}=10^{3}$. Within the limit $\lambda\gg1$, the
$|g_{1}\rangle\leftrightarrow|e_{1}\rangle\leftrightarrow|e_{2}\rangle\leftrightarrow|g_{2}\rangle$
three-photon Raman transition off-resonantly mediated by means of
the symmetric mode $a_{+}$ could be neglected, yielding
$r\rightarrow \sqrt{\left(\beta/\alpha+1\right)/2}$; hence,
according to Eqs.~(\ref{eq:Toffsuccessprob}) and
(\ref{eq:Tofferror}), the gate error is limited by an upper bound
[see Fig. {\ref{Figure4}(b)], along with the corresponding success
probability also upper bounded [see Fig. {\ref{Figure3}(b)].

\section{Heralded controlled-Z gate}

\label{se:CZgates} Let us now consider the heralded
near-deterministic realization of the two-qubit controlled-Z (CZ)
gate. This gate is also known as the two-qubit controlled-sign
gate, controlled-phase-flip gate, or controlled-phase gate.
Specifically, the action of the CZ gate in our system can be
explained as follows: Conditioned on the detection of the quartit
in its ground state $|g_{1}\rangle$, the CZ gate flips the phase
of the state $\ket{11}$ of an arbitrary two-qubit pure state
$\ket{\psi}=c_0\ket{00}+c_1\ket{01}+c_1\ket{01}+c_1\ket{11}$ or
any mixture of such states, where $c_k$ are the complex normalized
amplitudes and the qubit states correspond to the two
lowest-energy levels of the two qutrits in the resonator $B$.

It follows from Eq.~(\ref{eq:qubitdmatrix}) that, in order to
completely remove the gate error, the decay rate $\Gamma_{n}$
needs to be independent of the qutrits. To this end, we retune the
detunings $\delta$ and $\Delta_{e,2}$ to be
\begin{align}
\label{eq:CZlittledelta}
\frac{\delta}{\gamma}&=\frac{1}{2\left(2D+G^{-1}\right)},\\
\label{eq:CZDeltae2} \frac{\Delta_{e,2}}{\gamma}&=\alpha
C_{B}\left(D+G^{-1}\right),
\end{align}
where $D=\sqrt{\left[1/G^2+\beta/\left(\alpha
C_{B}\right)\right]/2}$. Thus, in the limit
$\left\{G,C_{B}\right\}\gg1$, retuning $\delta$ and $\Delta_{e,2}$
as above yields an $n$-independent decay rate
\begin{equation}
\Gamma_{n}=\Gamma,
\end{equation}
for all $n$, while Eq.~(\ref{eq:ToffGamma}) is valid only for $n=0,1$. The corresponding energy
shifts are given by
\begin{equation}
\Delta_{0}=-\frac{\Gamma D}{2},
\end{equation}
and
\begin{figure}[!ht]
\begin{center}
\includegraphics[width=8.5cm,angle=0]{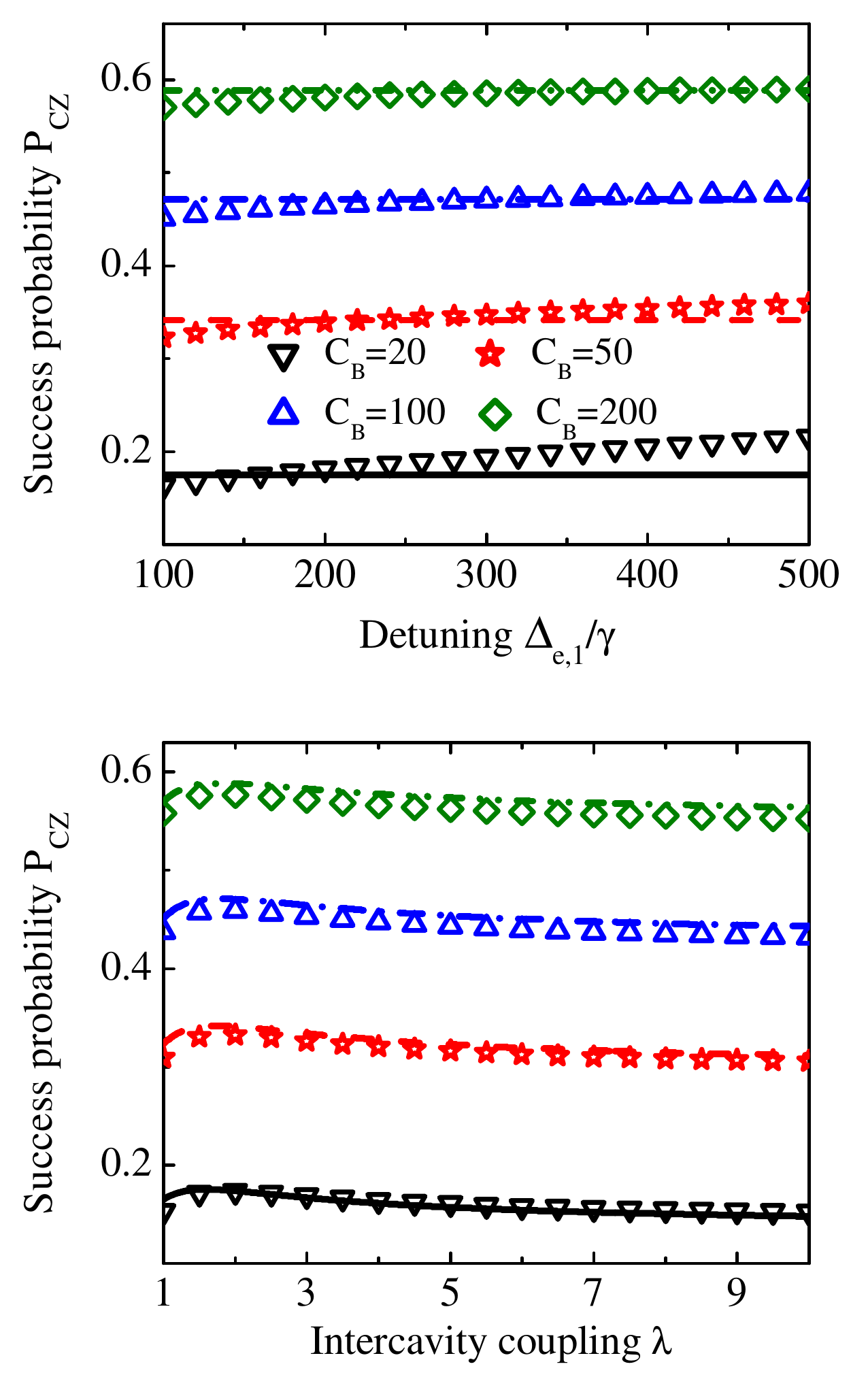}
\caption{(Color online) Numerical simulations (marked by symbols)
for the success probability, $P_{\text{CZ}}$, of the heralded CZ
gate, and for the cooperativity $C_{B}=20$ (black down-triangles),
$50$ (red stars), $100$ (blue up-triangles), and $200$ (olive
diamonds). The success probability is displayed versus the
detuning $\Delta_{e,1}$ given in terms of the overall decay rate
$\gamma$ in (a), as well as versus the inter-resonator coupling
strength $\lambda$ in (b). For comparison, we also plot the
analytical success probability (curves), and find that the
analytical results are in good agreement with the exact numerics.
Here, we have assumed $\lambda=1.84$ in (a) and
$\Delta_{e,1}=150\gamma$ in (b). In both panels we have set that
\new{$\gamma_{g,1}=\gamma_{g,2}=\gamma$}, $\gamma_{0}=\gamma_{1}=0.5\gamma$, $\kappa=10\gamma$,
\new{$g_{A}=g_{B}$, $C_{B}=g_{B}^{2}/\left(\kappa\gamma\right)$, $\lambda=J/\left(\kappa\sqrt{C_{B}}\right)$,}
$\alpha=\beta=1$, $\Omega_{1}=\Delta_{e,1}/(10C_{B}^{1/4})$, and
$\Omega_{2}=4\gamma C_{B}^{1/4}$.}\label{Figure5}
\end{center}
\end{figure}
\begin{equation}
\Delta_{n>0}=-\frac{\widetilde{\Omega}^{2}}{2\gamma}\frac{n\left(2D+
1/G\right)}{\alpha C_{B}\left(4nD^2+2nD/G+1/C_{B}\right)}.
\end{equation}
Subsequently, according to Eq.~(\ref{eq:qubitdmatrix}), the
conditional density operator of the qutrits becomes
\begin{equation}\label{eq:CZqubitdmatrix}
\rho_{\text{qutrit}}\left(t\right)=\sum_{n,n'=0}^{N}e^{-i\left(\Delta_{n}-\Delta_{n'}\right)t}\\
\mathcal{P}_{n}\left[\langle
g_{1}|\rho\left(0\right)|g_{1}\rangle\right]\mathcal{P}_{n'},
\end{equation}
where the decay factor, $\exp\left(-\Gamma t\right)$, has been
eliminated through a measurement conditional on the quartit atom
being detected in the state $|g_{1}\rangle$. Together with
single-qubit operations,  we can utilize the dynamics described by
the $\rho_{\text{qutrit}}\left(t\right)$ of Eq.
(\ref{eq:CZqubitdmatrix}) to implement a heralded CZ gate for
$N=2$.

For this purpose, the duration of the driving pulse is chosen to
be
\begin{equation}
t_{\text{CZ}}=\pi|\Delta_{2}-2\Delta_{1}+\Delta_{0}|^{-1},
\end{equation}
and the unitary operation on each qubit, applied after the pulse,
has the following action
\begin{equation}
\mathcal{U}|0\rangle=e^{i\Delta_{0}t_{\text{CZ}}/2}|0\rangle,
\quad
\mathcal{U}|1\rangle=e^{i\left(2\Delta_{1}-\Delta_{0}\right)t_{\text{CZ}}/2}|1\rangle,
\end{equation}
so as to ensure the right phase evolution. The resulting gate is
either to flip the phase of the qubit state $|11\rangle$, or to
leave the otherwise qubit states unchanged. The associated success
probability for any initial pure state is
\begin{equation}
P_{\text{CZ}}=\exp\left(-\Gamma t_{\text{CZ}}\right),
\end{equation}
which can, as long as $\left\{G,C_{B}\right\}\gg1$, be
approximated as
\begin{equation}
P_{\text{CZ}}=1-Z_{p}\frac{\pi}{\sqrt{C_{B}}},
\end{equation}
with a scaling factor
\begin{equation}
Z_{p}=2d+\frac{3}{2\left(2d+1/\lambda\right)}+\frac{1}{4d\left(2d+1/\lambda\right)^{2}},
\label{Z_P}
\end{equation}
where $d=\sqrt{\left(1/\lambda^2+\beta/\alpha\right)/2}$. If
assuming that the desired state is $|\psi\rangle$, we can
calculate the conditional fidelity for this gate via
\begin{equation} F_{\text{CZ}}=\langle
\psi|\left(\mathcal{U}\otimes\mathcal{U}\right)\rho_{\text{qutrit}}
\left(t_{\text{CZ}}\right)\left(\mathcal{U}\otimes\mathcal{U}\right)^{\dag}|\psi\rangle,
\end{equation}
and then can directly find $F_{\text{CZ}}=1$. This implies that
with the single-qubit operation, we achieve a more significant
improvement than that shown in Eq.~(\ref{eq:Tofferrorscaling}),
and, thus, by decreasing $\Omega_{1}$ and increasing
$\Delta_{e,1}$, an arbitrarily small gate error can even be
achievable.

In order to confirm the heralded CZ gate, we now perform numerical
simulations exactly using, instead of the effective master
equation in Eq.~(\ref{eq:effmastereq}), the full zero-temperature
master equation, given by Eq.~(\ref{ME}), for the density operator
$\rho_{T}\left(t\right)$ of the total system initially in
\begin{equation}
|\Psi\rangle_{\text{ini}}=|\Phi\rangle_{\text{ini}}\otimes|\text{vac}\rangle,
\end{equation}
with $|\text{vac}\rangle$ being the vacuum state of the coupled
resonators and $|\Phi\rangle_{\text{ini}}$ is given in Eq.
(\ref{eq:atomicinitialstate}). The numerical simulations calculate
the success probability $P_{\text{CZ}}$, the conditional density
operator $\rho_{\text{qutrit}}\left(t_{\text{CZ}}\right)$ and
fidelity $F_{\text{CZ}}$ using the expressions below:
\begin{equation}\label{eq:numsuprobg}
P_{\text{CZ}}=\sum_{n=0,1,2}\text{Tr}\left[\left(|g_{1}\rangle\langle
g_{1}|\otimes\mathcal{P}_{n}\otimes\mathcal{I}\right)\rho_{T}\left(t_{\text{CZ}}\right)\right],
\end{equation}
\begin{equation}\label{eq:numcondidensoperator}
\rho_{\text{qutrit}}\left(t_{CZ}\right)=\frac{1}{P_{\text{CZ}}}\text{Tr}_{\text{cav}}\left[\langle
g_{1}|\rho_{T}\left(t_{\text{CZ}}\right)|g_{1}\rangle\right],
\end{equation}
\begin{figure}[!ht]
\begin{center}
\includegraphics[width=8.5cm,angle=0]{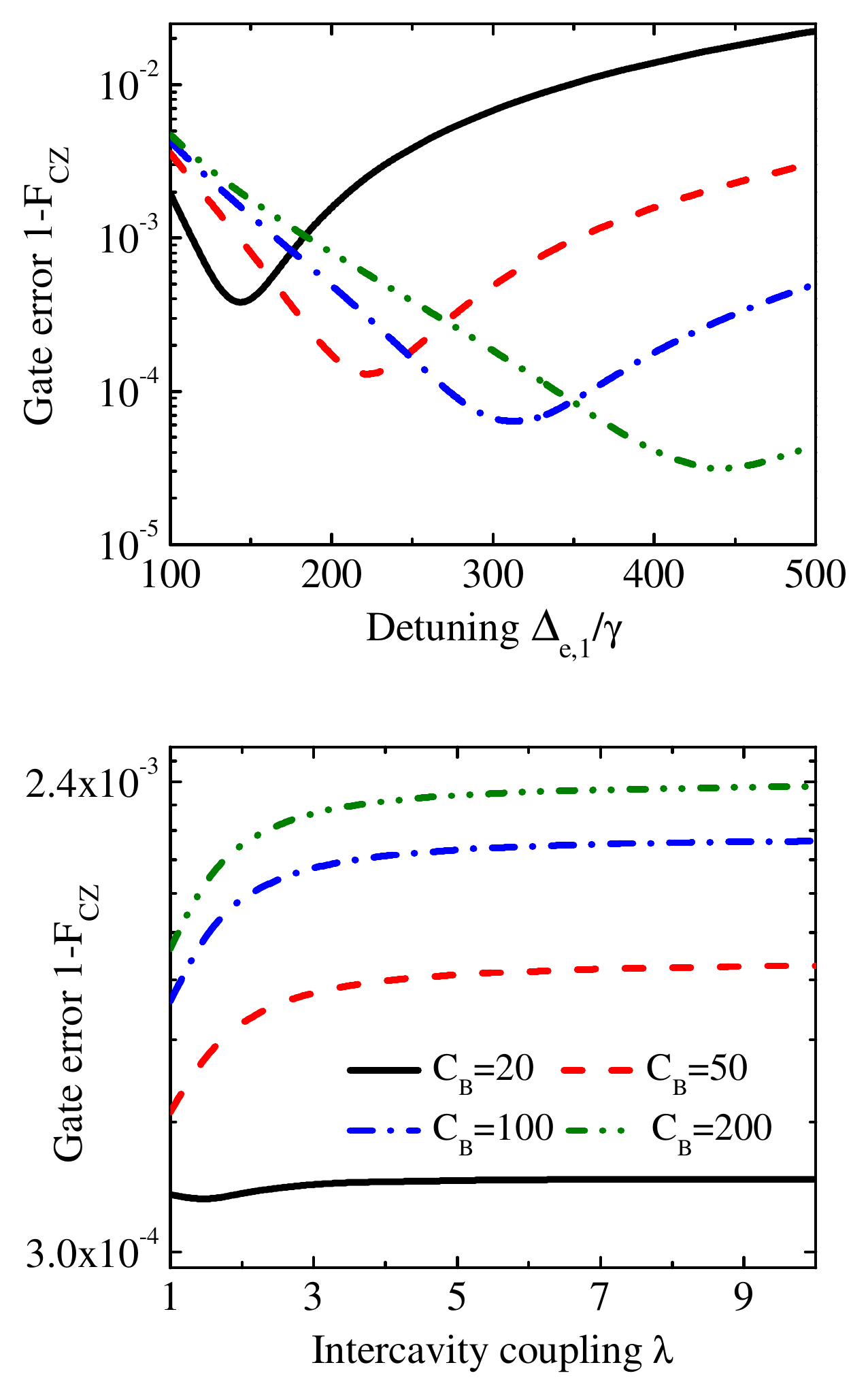}
\caption{(Color online) Numerical simulations for the gate error,
$1-F_{\text{CZ}}$, of the heralded CZ gate, and for the
cooperativity $C_{B}=$20, 50, 100, and 200. Upon setting the same
parameters as used in Fig.~\ref{Figure5}, the gate error is
displayed versus the detuning $\Delta_{e,1}/\gamma$ in (a), as
well as versus the inter-resonator coupling strength $\lambda$ in
(b). There is a one-to-one correspondence between the gate error
here and the numerically calculated success probability
$P_{\text{CZ}}$ in Fig.~\ref{Figure5}.}\label{Figure6}
\end{center}
\end{figure}
\begin{equation}\label{eq:numcondifidelity}
F_{\text{CZ}}=\langle\psi|\left(\mathcal{U}\otimes\mathcal{U}\right)
\rho_{\text{qutrit}}\left(t_{\text{CZ}}\right)\left(\mathcal{U}\otimes\mathcal{U}\right)^{\dag}
|\psi\rangle,
\end{equation}
where $\text{Tr}$ and $\text{Tr}_{\text{cav}}$ are trace
operations over the total system and the resonators, respectively,
and $\mathcal{I}$ is an identity operator related to the two
resonators. For the simulations, we assume that
$\gamma_{g,1}=\gamma_{g,2}$, $\gamma_{0}=\gamma_{1}=0.5\gamma$,
and $\alpha=\beta=1$; moreover, we take $\kappa=10\gamma$,
$\Omega_{1}=\Delta_{e,1}/(10C_{B}^{1/4})$,
$\Omega_{2}=4\gamma C_{B}^{1/4}$, and determine the detunings
$\delta$ and $\Delta_{e,2}$ according to
Eqs.~(\ref{eq:CZlittledelta}) and~(\ref{eq:CZDeltae2}),
respectively. Then, we calculate the success probability
$P_{\text{CZ}}$ and the gate error $1-F_{\text{CZ}}$, as a
function of either the detuning $\Delta_{e,1}$ or the
inter-resonator coupling $J$, for different
cooperativities~\cite{qutip1,qutip2}. The numerical results
(marked by symbols) are plotted in Fig.~\ref{Figure5} for the
success probability and in Fig.~\ref{Figure6} for the gate error.
The analytical success probability, which we use curves to
represent, is also plotted in Fig.~\ref{Figure5}, and shows good
agreement with the exact results. In Fig.~\ref{Figure6}(a) we find
that as the detuning $\Delta_{e,1}$ increases, the gate error
first decreases and then increases. This is because, in addition
to suppressing the error from the
$|e_{1}\rangle\rightarrow|g_{1}\rangle$ decay (see the Appendix),
such an increase in $\Delta_{e,1}$, however, increases the driving
strength $\Omega_{1}=\Delta_{e,1}/(10C_{B}^{1/4})$, and
hence the non-adiabatic error. There is a trade-off between the
two errors. Again within the limit $\lambda\gg1$, the off-resonant
Raman transition could be removed as before, yielding
$d\rightarrow \sqrt{\beta/\left(2\alpha\right)}$. As a result, one
finds that the scaling factor, given by Eq.~(\ref{Z_P}), is equal
to
\begin{equation}
  Z_{p}\rightarrow2d+\frac{3}{4d}+\frac{1}{16d^{3}}.
 \label{Z_P2}
\end{equation}
Hence, when $\lambda$ is sufficiently large, both the success
probability and the gate error will be independent of $\lambda$,
illustrated in Figs.~\ref{Figure5}(b) and~\ref{Figure6}(b). It
should be noted that to calculate the success probability in Fig.
\ref{Figure5}(a), and the gate error in Fig.~\ref{Figure6}(a), we
have chosen $\lambda=1.84$, because this value can lead to the
shortest driving pulse for the two-qubit gate if $\alpha=\beta$.


Finally, we note that the energy shifts, $\Delta_{n}$, for
both Toffoli-like and CZ gates involve only the ground states,
because the effective master equation~(\ref{eq:effmastereq}) is
obtained by adiabatically eliminating the excited states.
According to the effective Hamiltonian of Eq.~(\ref{eq:effH}), a
quantum state, with $n$, of the qutrit atoms has an energy shift,
$\Delta_{n}$. Consequently, a dynamical phase,
$\exp(-i\Delta_{n}t)$, can be imposed by choosing the appropriate
evolution times for a given quantum state. Thus, the quantum
gates, including the multi-qubit Toffoli-like (see Table II) and
two-qubit CZ gates (Table III), are realized by choosing
appropriate evolution times.

\section{conclusions}
We have described a method for performing a heralded
near-deterministic controlled phase gates in two distant
resonators in the presence of decoherence, including the
two-qubit controlled-Z (CZ) gate and its multi-qubit-controlled
generalization known, which can be referred to as a Toffoli-like
gate. Our proposal is a generalization of the single-resonator
method introduced by Borregaard {\it et al.}~\cite{Dgate1}.
The method in our paper uses an auxiliary microwave-driven four-level
atom (quartit) inside one resonator to serve as both an
intra-resonator photon source and a detector, and, thus, to
control and also herald the gates on atomic qutrits inside the
other resonator. In addition to the quantum gate fidelity scalings, which are
superior to traditional cavity-assisted deterministic gates, this
method demonstrates a macroscopically-distant herald for
controlled quantum gates, and at the same time avoids the
difficulty in individually addressing a microwave-driven atom.
We note that the original method of Ref.~\cite{Dgate1} is based on
the heralding and operational atoms coupled to the same resonator
mode acting as a cavity bus. Here, we are not applying this cavity
bus.

The operator formalism used in our paper to calculate
the effective Hamiltonian [i.e.,
Eq.~(\ref{eq:formalism-for-effHam})] and the effective Lindblad
operators [i.e., Eq.~(\ref{eq:formalism-for-effLind})] follows the
approach in Ref.~\cite{effmaseq}. The dynamical behavior of the
system can be fairly-well approximated by the effective master
equation. For example, the results obtained respectively by using
the full and effective master equations in Figs. 4 and 6 in
Ref.~\cite{effmaseq}, or in Figs.~\ref{Figure5} and~\ref{Figure6}
in our paper, are in good agreement. The physics of this
approximation can be understood by the simple qutrit atom in
Fig.~\ref{Figure2}. A direct picture is that the three-level atom
in state $|0\rangle$ is first excited to the state $|2\rangle$,
then can go to the state $|1\rangle$ by an atomic decay, or can
go back to the state $|0\rangle$ also by another atomic decay. If
starting in the state $|1\rangle$, the atom has a similar behavior.
Roughly speaking, there exists an indirect interaction between the
states $|0\rangle$ and $|1\rangle$, as well as the direct interaction
between the states $|0\rangle$, $|1\rangle$, and the state
$|2\rangle$, mediated by the atomic decays. Thus, after
adiabatically eliminating the state $|2\rangle$, the energy shifts
of the states $|0\rangle$, $|1\rangle$, and even a direct
interaction between them, would be induced by the atomic decays.
Upon combining the above processes, mediated by the atom decays
with the coherent drives and other interactions, the effective
master equation was obtained here in analogy to that in
Ref.~\cite{effmaseq}.

We have also proposed a circuit-QED system with superconducting
qutrits and quartits implementing the proposed protocol as shown
in Fig.~\ref{Figure1}. Our circuit includes two coupled
transmission-line resonators, which are linearly coupled via a
SQUID. These resonators can be coupled to qudits via a
capacitance. For example, we assumed a phase quartit coupled to
one of the resonators serving a herald, and $N$ identical flux
qutrits coupled to the other resonator for performing the
controlled phase gates.  Typically, the length of a
transmission-line resonator can be of up to the order of
cm~\cite{rSillanpaa07}, but the size of a superconducting atoms is
of $\sim \mu$m~\cite{rXiang13}.

\new{We assume realistic parameters from experiments with
superconducting quantum circuits~\cite{niemczyk2010circuit,rSillanpaa07}. Specifically: $\gamma/2\pi=10$~MHZ, $\kappa/2\pi=6$~MHz, $C_{A}=C_{B}=170$ ($\alpha=\beta=1$), $\Delta_{e,1}/2\pi=420$~MHz, $\Omega_{1}/2\pi=70$~MHz, $\lambda=1.84$, $J/2\pi=144$~MHz, and $\Omega_{2}/2\pi=87$~MHz. The implementation of the CZ gate with these parameters can result in a success probability of $\sim 0.55$, a gate error of $\sim 0.006$, and a gate time of $\sim6~\mu $s. These gate performance parameters are quite good for a coupled-resonator system. Indeed by increasing the capacitance of the capacitor between the TLRs, it is possible to achieve a stronger $J$~\cite{Fitzpatrick2016observation} and, thus, a smaller gate error and a shorter gate time. Furthermore, the decoherence time $T_{2}$ of a flux qubit can be improved to $\sim85~\mu$s~\cite{Yan15theflux}, which is much larger than $6~\mu$s. This justifies neglecting such decoherence effect on the flux qutrits. But for a phase qubit, $T_{2}$ reaches $\sim1~\mu$s, which is smaller than $6~\mu$s~\cite{Martinis09superconducting}. Nevertheless, the phase quartit in our protocol works only as an auxiliary atom to herald the quantum gate by measuring the $|g_{1}\rangle$ population on which the decoherence, quantified $T_{2}$,  has no effect. Hence, such
decoherence of both phase quartit and flux qutrits can be effectively neglected in our protocol.}

\new{Another possible implementation can be based on
ultracold atoms coupled to nanoscale optical cavities. In this
situation, the atoms $^{87}\text{Rb}$ can be used for both quartit and qutrit atoms~\cite{Dgate1,exp:atom1,exp:atom2}.
Furthermore, nanoscale optical cavities have been realized by
the use of defects in a two dimensional photonic crystal, and one
can place such nanocavities very close to each other to
directly couple them by evanescent
fields~\cite{exp:distantcavity1}, or one can use a common
waveguide (a quantum bus) to indirectly couple such
cavities~\cite{exp:distantcavity2}}.

Although we have chosen to discuss the specific case of two
coupled resonators, this description may be extended to a
coupled-cavity array~\cite{CCA1,CCA2,CCA3,CCA4,CCA5}. Hence, it
would enable applications such as scalable quantum information
processing and long-distance quantum communication.

\section{acknowledgments}

W.Q. is supported by the Basic Research Fund of the Beijing
Institute of Technology under Grant No. 20141842005. X.W. is
supported by the China Scholarship Council (Grant No.
201506280142). A.M. and F.N. acknowledge the support of a grant
from the John Templeton Foundation. F.N. was also partially
supported by the RIKEN iTHES Project, the MURI Center for Dynamic
Magneto-Optics via the AFOSR award number FA9550-14-1-0040, the
IMPACT program of JST, CREST grant No. JPMJCR1676, a Grant-in-Aid for Scientific
Research (A), the Japan Society for the Promotion of Science (KAKENHI), JSPS-RFBR grant No. 17-52-50023.

\appendix* 
\section{\noindent Derivation of effective master equation}\vspace{6pt}
\label{appendix} In this Appendix, we will derive the effective
master equation by using the method in Ref.~\cite{effmaseq}. We
start with the total Hamiltonian $H_{T}$ in the main text. Upon
introducing the symmetric and antisymmetric modes,
$a_{\pm}=\left(a_{A}\pm a_{B}\right)/\sqrt{2}$, of the coupled
resonators, and then switching into a rotating frame with respect
to
\begin{align}
H_{\text{rot}}=&\sum_{k=1}^{N}\left[\left(\omega_{c}+\omega_{1}-J\right)|2\rangle_{k}\langle 2|+\sum_{z=0,1}\omega_{z}|z\rangle_{k}\langle z|\right]\nonumber\\
&+\bar{\omega}|e_{1}\rangle\langle e_{1}|+\left(\bar{\omega}+\omega_{m,2}\right)|e_{2}\rangle\langle e_{2}|\nonumber\\
&+\sum_{x=1,2}\omega_{g,x}|g_{x}\rangle\langle
g_{x}|+\left(\omega_{c}-J\right)\left(a_{+}^{\dag}a_{+}+a_{-}^{\dag}a_{-}\right),
\end{align}
where $\bar{\omega}=\omega_{m,1}+\omega_{g,1}$, the total
Hamiltonian is transformed to
\begin{equation}
H_{T}=H_{e}+V+V^{\dag},
\end{equation}
as given in the main text. With the Lindblad operators in Eq.
(\ref{eq:lindbladoperators}), we obtain the no-jump Hamiltonian of
the form
\begin{align}
H_{\text{NH}}
=&\sum_{k=1}^{N}\left\{\bar{\delta}|2\rangle_{k}\langle 2|+\frac{g_{B}}{\sqrt{2}}\left[\left(a_{+}-a_{-}\right)|2\rangle_{k}\langle1|+\text{H.c.}\right]\right\}\nonumber\\
&+\bar{\Delta}_{e,1}|e_{1}\rangle\langle e_{1}|+\bar{\Delta}_{e,2}|e_{2}\rangle\langle e_{2}|+\bar{J}a_{+}^{\dag}a_{+}\nonumber\\
&-\frac{i\kappa}{2}a_{-}^{\dag}a_{-}+\frac{g_{A}}{\sqrt{2}}\left[\left(a_{+}+a_{-}\right)|e_{2}\rangle\langle g_{2}|+\text{H.c.}\right]\nonumber\\
&+\frac{\Omega_{2}}{2}\left(|e_{2}\rangle\langle
e_{1}|+\text{H.c.}\right),
\end{align}
where $\bar{\Delta}_{e,1}=\Delta_{e,1}-i\gamma_{g,1}/2$,
$\bar{\Delta}_{e,2}=\Delta_{e,2}-i\gamma_{g,2}/2$,
$\bar{\delta}=\delta-i\gamma/2$, and $\bar{J}=2J-i\kappa/2$.
Following the formalism in Ref.~\cite{effmaseq}, the effective
Hamiltonian and Lindblad operators are given, respectively, by
\begin{equation}
H_{\text{eff}}=\frac{1}{2}V^{\dag}\left[H_{\text{NH}}^{-1}+\left(H_{\text{NH}}^{-1}\right)
^{\dag}\right]V,
\end{equation}
\begin{equation}
L_{\text{eff}}^{j}=L_{j}H_{\text{NH}}^{-1}V.
\end{equation}
To proceed, we work within the single-excitation subspace and,
after a straightforward calculation, obtain
\begin{align}
\Delta_{n}=-\frac{\Omega_{1}}{2\sqrt{\gamma_{g,1}}}\text{Re}\left(r_{g_{1},n}\right),
\end{align}
\begin{equation}
r_{+,n}=\frac{\Omega_{1}\widetilde{\Omega}_{m}\sqrt{C_{A}}}{4\widetilde{J}\mathcal{X}_{n}\sqrt{2\gamma}}\left(i\widetilde{\delta}+2nC_{B}\right),
\end{equation}
\begin{equation}
r_{-,n}=-\frac{\Omega_{1}\widetilde{\Omega}_{m}\sqrt{C_{A}}}{2\mathcal{X}_{n}\sqrt{2\gamma}}\left(\widetilde{\delta}-nC_{B}/\widetilde{J}\right),
\end{equation}
\begin{align}
r_{g_{1},n}=&\frac{\Omega_{1}\sqrt{\gamma_{g,1}}}{2\gamma \mathcal{X}_{n}}\bigg[i\widetilde{\delta}\widetilde{\Delta}_{e,2}+\left(C_{A}\widetilde{\delta}+nC_{B}\widetilde{\Delta}_{e,2}\right)\nonumber\\
&\times\left(1-i/2\widetilde{J}\right)-2nC_{A}C_{B}/\widetilde{J}\bigg],
\end{align}
\begin{equation}
r_{g_{2},n}=-\frac{\Omega_{1}\widetilde{\Omega}_{m}\sqrt{\gamma_{g,2}}}{4\gamma
\mathcal{X}_{n}}\left\{i\widetilde{\delta}+nC_{B}\left[1-i/(2\widetilde{J})\right]\right\},
\end{equation}
\begin{equation}
r_{k,n}=-\frac{\Omega_{1}\widetilde{\Omega}_{m}\sqrt{\gamma_{k}C_{A}C_{B}}}{4\gamma
\mathcal{X}_{n}}\left[1+i/(2\widetilde{J})\right],
\end{equation}
where  $C_{A}=g_{A}^{2}/\left(\kappa\gamma\right)$,
$\widetilde{\Delta}_{e,1}=\Delta_{e,1}/\gamma-i\gamma_{g,1}/\left(2\gamma\right)$,
$\widetilde{J}=2J/\kappa-i/2$,
$\widetilde{\Omega}_{2}=\Omega_{2}/\gamma$,
$Z=\widetilde{\Delta}_{e,1}\widetilde{\Delta}_{e,2}-\left(\widetilde{\Omega}_{m}/2\right)^{2}$,
$\mathcal{X}_{n}=iZ\widetilde{\delta}+\left(C_{A}\widetilde{\delta}\widetilde{\Delta}_{e,2}+nC_{B}Z\right)\left[1-i/(2\widetilde{J})\right]
-2nC_{A}C_{B}\widetilde{\Delta}_{e,1}/\widetilde{J}$, and $k=0,1$.
In the limit $\Omega_{2}\ll \Delta_{e,1}$, we can make a Taylor
expansion around $\Omega_{2}/\Delta_{e,1}=0$, yielding
\begin{equation}\label{eqa:deltan02}
\Delta_{n}=-\frac{\Omega_{1}^{2}}{4\Delta_{e,1}}
-\frac{\widetilde{\Omega}^{2}}{4\gamma}
\text{Re}\left\{\frac{1}{\mathcal{Y}_{n}}\left\{i\widetilde{\delta}+nC_{B}
\left[1-\frac{i}{2\widetilde{J}}\right]\right\}\right\},
\end{equation}
\begin{equation}\label{eqa:r+n02}
r_{+,n}=\frac{\widetilde{\Omega}\sqrt{C_{A}}}{2\widetilde{J}\mathcal{Y}_{n}\sqrt{2\gamma}}\left(i\widetilde{\delta}+2nC_{B}\right),
\end{equation}
\begin{equation}\label{eqa:r-n02}
r_{-,n}=-\frac{\widetilde{\Omega}\sqrt{C_{A}}}{\mathcal{Y}_{n}\sqrt{2\gamma}}\left(i\widetilde{\delta}-nC_{B}/\widetilde{J}\right),
\end{equation}

\begin{equation}\label{eqa:rgn02}
r_{g_{1},n}=\frac{\Omega_{1}\sqrt{\gamma_{g,1}}}{2\Delta_{e,1}}
+\frac{\widetilde{\Omega}\sqrt{\widetilde{\gamma}_{g,1}}}{2\gamma
\mathcal{Y}_{n}}\left\{i\widetilde{\delta}+nC_{B}\left[1-\frac{i}{2\widetilde{J}}\right]\right\},
\end{equation}
\begin{equation}
r_{g_{2},n}=-\frac{\widetilde{\Omega}\sqrt{\gamma_{g,2}}}{2\gamma
\mathcal{Y}_{n}}\left\{i\widetilde{\delta}+nC_{B}\left[1-i/(2\widetilde{J})\right]\right\},
\end{equation}
\begin{equation}\label{eqa:rxn02}
r_{k,n}=-\frac{\widetilde{\Omega}\sqrt{\gamma_{k}C_{A}C_{B}}}{2\gamma
\mathcal{Y}_{n}}\left[1+i/(2\widetilde{J})\right],
\end{equation}
where
$\widetilde{\Omega}=\Omega_{1}\Omega_{2}/\left(2\Delta_{e,1}\right)$,
$\widetilde{\gamma}_{g,1}=\gamma_{g,1}\Omega^{2}_{2}/\left(4\Delta_{e,1}^{2}\right)$,
and
$\mathcal{Y}_{n}=i\widetilde{\delta}\widetilde{\Delta}_{e,2}+\left(C_{A}
\widetilde{\delta}+nC_{B}\widetilde{\Delta}_{e,2}\right)
\left[1-i/(2\widetilde{J})\right]-2nC_{A}C_{B}/\widetilde{J}$. In
Eq.~(\ref{eqa:deltan02}), the term,
$-\Omega_{1}^{2}/\left(4\Delta_{e,1}\right)$, of $\Delta_{n}$ can
be removed because it has no effect on the phase gates. Meanwhile,
in Eq. (\ref{eqa:rgn02}), we find that the
$|e_{1}\rangle\rightarrow|g_{1}\rangle$ decay is suppressed by
increasing $\Delta_{e,1}$, such that the second term of
$r_{g_{1},n}$ can be removed as long as
$\widetilde{\gamma}_{g,1}\ll1$. Thus,
Eqs.~(\ref{eqa:deltan02})--(\ref{eqa:rxn02}), under the assumption
that $\kappa\ll J$, reduce to Eqs. (\ref{eq:deltan_a})
and~(\ref{eq:effectivelindbladoperators_a}) in the main text.

%

\end{document}